# A Secure Multi-Party Computation Protocol for Malicious Computation Prevention for preserving privacy during Data Mining

Dr. Durgesh Kumar Mishra
Professor (CSE) and Dean (R&D),
Acropolis Institute of Technology & Research,
Indore, MP, India
mishra_research@rediffmail.com

Neha Koria, Nikhil Kapoor, Ravish Bahety
Computer Science Dept. (R&D)
Acropolis Institute of Technology & Research,
Indore, MP, India
koria.neha@gmail.com, nikhilkapoor01@gmail.com,
ravish.bahety@gmail.com

ABSTRACT- **Secure Multi-Party Computation (SMC) allows parties with similar background to compute results upon their private data, minimizing the threat of disclosure. The exponential increase in sensitive data that needs to be passed upon networked computers and the stupendous growth of internet has precipitated vast opportunities for cooperative computation, where parties come together to facilitate computations and draw out conclusions that are mutually beneficial; at the same time aspiring to keep their private data secure. These computations are generally required to be done between competitors, who are obviously weary of each-others intentions. SMC caters not only to the needs of such parties but also provides plausible solutions to individual organizations for problems like privacy-preserving database query, privacy-preserving scientific computations, privacy-preserving intrusion detection and privacy-preserving data mining. This paper is an extension to a previously proposed protocol Encrytpo_Random, which presented a plain sailing yet effective approach to SMC and also put forward an aptly crafted architecture, whereby such an efficient protocol, involving the parties that have come forward for joint-computations and the third party who undertakes such computations, can be developed. Through this extended work an attempt has been made to further strengthen the existing protocol thus paving the way for a more secure multi-party computational process.**

## 1. INTRODUCTION

The SMC has been a problem that has attracted the attention of scholars and the industry for quite some time. Although a vast amount of work has been done upon the subject, the perpetual implementation of the endeavors has only yielded a perennial hornet's nest. Having said that, it should be acknowledged that to compute results upon data whose source is not known is not child's play; and the works undertaken until now have served a great purpose in enlightening the industry of the subtleties of this so-called SMC problem.

Thus motivated with the intention of solving this SMC problem we proposed a new protocol *Encrytpo_Random* through which we had put forward what we perceived, to be the most appropriate and seemingly plausible solution to the SMC conundrum. The methodology followed was quite elementary yet very comprehensible. *Encrytpo_Random* worked on a two layer basis; it consisted of the parties (1st layer) who aspire to draw out a result collectively and being apprehensive of each-others intentions appoint an assumedly unbiased third party (2nd layer) to carry out the computation and announce the result.

In *Extended Encrytpo_Random* the domain of the 2nd layer has been extended from a single third-party to multiple third-parties, from whom a single entity is chosen at run time and given the responsibility of performing the required computation. A proposal sounds overtly hyperbolic without a thorough layout of the architecture to aptly implement it. Thus, here we also present a meticulously worked-out architecture to realize the protocols and also to showcase and answer the pertinent queries that are bound to arise in the minds of the audience.

The modus-operandi of the protocol deters the bodies involved to exhibit any malicious conduct by presenting thoroughly planned impediments in the path of the transfer of data among themselves. The security of information of the parties is of utmost importance in any approach seeking to solve the SMC enigma. In our protocols we have taken adequate precautions so as to guarantee the security of data of the involved parties. Instead of sending the entire data blocks the parties break them into packets and randomly distribute amongst themselves, for a stipulated number of times. Provisions have been done so as to ensure that the parties do not get to know whose data packets they are forwarding, and in stark contrast, the third party also doesn't have even a Lilliputian hint as to whose data packet a particular party is sending. This necessitates the need of a secure channel to transfer the data packets which have been dealt with in the deftly formed and apposite architecture. To further conceal the identity of the data packets we apply an encrypting function upon the data packets; these encrypting functions also reach to the third party through the same path and are used to decode the packets and rearrange them to form data blocks.

## 2. BACKGROUND

The SMC came to the fore-front as the Millionaires problem described by Yao in [8]. U.Maurer considered the general SMC protocols [7]. For specific tasks like online auctions, public voting or online updating of the data there exist very efficient and effective protocols. General SMC protocols are less effective than special purpose protocols. Maurer also defined the different types of security in databases [5]. Privacy preserving data mining using SMC has great importance and many applications have been developed [9, 14]. Du et al. reviewed the various industrial problems and listed them in [1, 3]. Some of the existing protocols are in the form of the circuit evaluation protocols, and encryption with homomorphism schemes.

The first general constant round protocol for secure two-party computations was given by Yao [16]. Yao's original protocol considered only the case of semi-honest parties. An extension to the case of malicious party was given by Lindell [17]. Goldreich et al. showed the existence of a secure solution of SMC problem [6]. The size of such protocols depend upon the number of parties involved in the computation process.

A new concept was put forward by D.K.Mishra and M.Chandwani [18] through their multi-layer protocols. Initially a two-layer protocol along with a tentative architecture for its implementation was proposed. This two-layer protocol was improvised by a three-layer protocol in which an anonymizer layer was added in between the participating parties and the third party. This new layer hid the information of the parties from the third party, who computes the data and provides the result. In the next paper [19], this three-layer protocol, was further extended into a four-layer protocol in which a packet layer was introduced. This new concept provided security to the data from most of the malicious activity, even if the third party is not a trusted one.

The magnanimity and the complexity of this protocol present an unusual paradox. On one hand owing to its compounded and uncanny nature, this protocol prevents most of the unscrupulous activities; but on the other hand, its intricate technicalities make it very difficult to actually implement it. Another disadvantage of this protocol is that, the anonymzier-layer has been assumed to be incorruptible; if in case it becomes malicious then many an information leak can occur.

To solve such problems we proposed a new protocol *Encrytpo_Random*[20], which involved only two layers: 1st Layer consisted of the parties that wish to compute results and the 2nd Layer made of the Trusted Third Party (TTP), that which computes the result for these parties. Here we are putting forward an extension with the aim of further strengthening the existing protocol. We consider the third party as a trusted one on account that it computes the results correctly.

## 3. INFORMAL DESCRIPTION

### 3.1 *Encrypto_Random*

The previous attempts done to solve the SMC problem were either too simple or gullible in their outlook that it was quite easy for the various parties involved in the process of computation to leak the data; or were too complicated to be realized in the real world.

Our protocols dare to rectify the fallacies of these preceding works by proposing a new scheme for solving the SMC problem. The main aspect of SMC is the computation of data in a secure and private manner. Thus the handling of the data is the major concern. Keeping this vital point of the SMC in mind, we put forward a simple but effective two-layer protocol. The layout of our protocol is straight forward, which makes is easy to implement; but we have put various checks at subsequent levels so as to ensure the security of the data of the incumbent parties. *Encrypto_Random* works on the simple technique that 'n' number of parties decide to coordinate, and thus need to put forward their data; obviously aspiring not to give undue advantage to their competitors by revealing their information. Thus they appoint a third party to compute upon their collective data and announce the result publicly.

*Encrypto_Random* exhibits two-layer architecture, activation of each layer being an alternate phenomenon. At the 1st layer the Parties (who wish to perform the computations) are active while at the 2nd layer the Third Party (who computes the result for the parties) is active. The parties break their data blocks into packets (number of packets formed by each party is fixed, as decided by the TTP). Then an encrypting function is used by each party to encode their data packets. This encrypting function is drawn out randomly by each party from a pool maintained by the Third Party. These encrypted packets are then randomly sent to other participating parties. This random distribution is synchronized, and is realized by using a random function which can be initiated by any of the participating parties at run time. This process of random dissemination of the data packets is carried out many times; which further guarantees the confidentiality of the data blocks. After a certain number of this parceling out, these data packets are then sent to the third party; who has the resources to store all the incoming data packets. The TTP then decodes the received data packets by using the pool of encrypting functions maintained by it, consequently re-arranging them into whole data blocks. Now the computations are carried out as per the collective requirements of the parties and the result is announced publicly.

### 3.2 *Extended Encrypto_Random*

*Encrypto_Random* gives an efficient and secure way of carrying out multi-party computations. Requisite measures were taken to check the flow of data. But still a certain amount of hesitancy creeps into the mind of the observer that too much of information is being leveraged upon the TTP. To take care of such doubts we here propose an extension to our existing work, wherein the computational layer consists of not one but multiple third parties. One out of these many TTP is chosen at run time, and all data packets are then forwarded by the parties on the 1st layer to this TTP. This TTP then rearranges the data packets into data blocks and consequentially calculates and announces the result. Choice of the TTP is done on a random basis and is undertaken by another randomization function, in-built in the architecture of the protocol.

As the TTP is not predetermined and is unknown until the time of the computation, this decreases the possibility of a joint malicious conduct by the parties and the TTP. Also, since most of the activities take place at run-time, it guarantees against any sort of unscrupulous activities (joint or individual) by the various entities as well. The encrypting functions are selected from a pool maintained by each TTP, thus the parties does not individually need to intimate the TTP as to which function they have used. The data packets are distributed randomly amongst the parties, that too a number of times, because as when the parties will forward the data it will be a combination of various packets and thus the identity is hidden.

Here special care has been taken to keep the process very straight-forward for the parties. The chosen TTP however has to undertake a lot of efforts to decrypt the data packets and to rearrange them back into whole data blocks. But this is just a small price to pay for maintaining the secrecy of information, which is of utmost importance for every organization.

### 4. ASSUMPTIONS

The following assumptions have been stipulated:
1. TTP computes the result from the data provided to it by the parties.
2. It will do so by using a function, for which it will bring into effect a function generator.
3. TTP has the ability to announce the result of the computation publicly, although this won't be desirable in most of the cases.
4. Each party having an input can communicate with a trusted network connection.
5. The communication channels used by the input providing parties to communicate with the third party are secure. This implies that no intruder can intercept the data transferred between them.
6. The function given to the domain of the parties are not same, they are different.
7. Minimum three parties should be involved in the SMC.
8. The number of packets generated by each party is the same, as decided by the TTP.
9. The packet sizes of the parties are equal, as decided by the TTP.
10. The TTP has the resources to store all the incoming data packets and the encoded functions.

### 5. ARCHITECTURE

#### 5.1 *Encrypto_Random:*

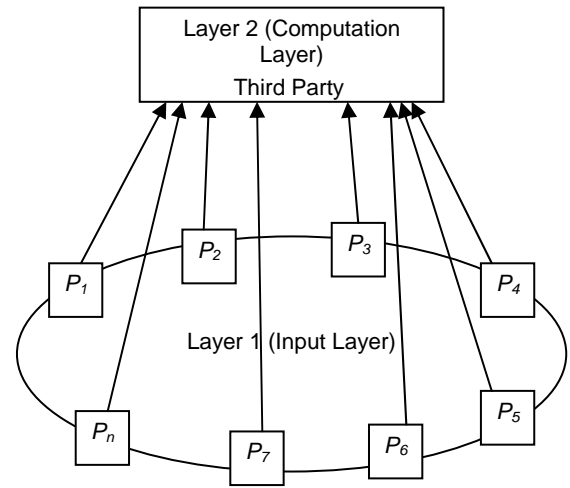

**Figure 1. The Existing Architecture**

Figure 1 depicts the simple architecture of *Encrypto_Random*. According to this the 1st layer i.e. The Input Layer comprise of all the parties that are involved in the computation process. Since all these parties are inter-connected, the data packets of the respective parties are randomly distributed amongst them, for a stipulated number of times. Thus when the parties forward the data packets, the Third Party receives the data packets that belong to some other party rather than the one who has forwarded them. This ensures that the identity of the parties is hidden in terms of which data packets belong to them.

The Third Party exists at the 2nd layer, i.e. The Computation Layer, where the computations are carried out. After receiving the data packets from the parties (at layer 1), the data packets are re-arranged by the Third Party using the Pool of Functions. Once the data packets are reassembled the computation is carried out and the result is obtained. It seems to be a simple architecture but provides various checks keeping in mind the fulfillment of the all the constraints of the SMC.

## 5.2 *Extended Encrypto_Random:*

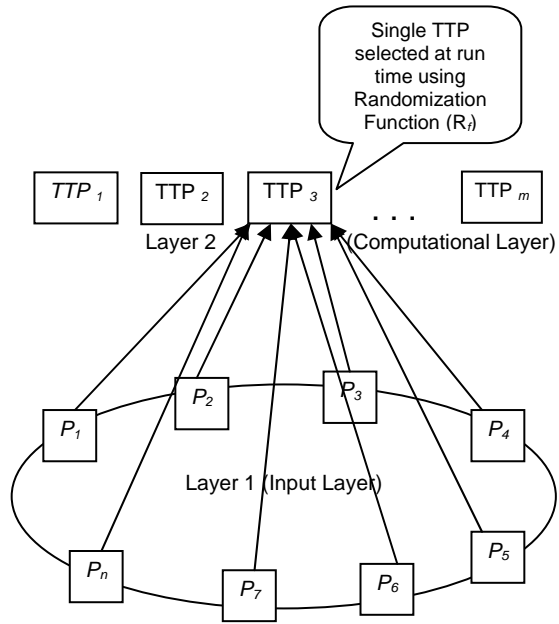

**Figure 2. The Proposed Architecture**

Figure 2 depicts the extended architecture of our protocol. Instead of a single TTP, the computational layer (2nd layer) consists of a pool of TTP. Each TTP has the same pool of functions, which were used previously by the parties (1st layer) to encrypt their data packets. One out these many TTP is chosen at run time and all data packets are forwarded to it making it responsible for the regrouping of data packets into data blocks and the subsequent required calculations.

## 6. FORMAL DESCRIPTION

*Extended Encrypto_Random*

The extended protocol, like its predecessor, is based on the simple pretext of a group of parties $(P_1, P_2...P_n)$ who assign the task of computation to a TTP. Each party $P_n$ breaks its data block into number of packets $(P_nK_r)$ and applies encrypting function $(f_1, f_2...f_n)$ chosen from the pool of functions $(D)$. This breaking of the data blocks into packets and the application of the encrypting functions is facilitated by the already existing provisions deftly incorporated in the architecture of the protocol. These encrypted packets $(S_{nr}= P_nK_r + V_rF_n)$ are randomly distributed among parties. A TTP is chosen at run time using randomization function $(R_f)$. All data is sent to this TTP, who then decodes the data packets, rearranges them into whole data blocks; computes and announces the result publicly.

**Algorithm** *Extended Encrypto_Random:*
1. Define $P_1, P_2, ..., P_n$ as parties;
2. Define $D$ as function pool;
3. Define $F_1, F_2, ..., F_n$ as encrypting functions;
4. For party $P_1$ to $P_n$ do
   begin
   Break data block into packets;
   $(P_nK_1, P_nK_2 ... P_nK_r)$
   /* Where $K_1, K_2 ... K_r$
   designate the packets of a party */
   Select function from pool $(D)$;
   Attach encrypting value of function to each packet;
   Compute $S_{nr} = P_nK_r + V_rF_n$;
   /* Where $V_1, V_2, ..., V_r$ are the values of function $F_n$*/
   end;
5. For $r=1$ to $s$, do
   begin
   Send $S_{nr}$ randomly to $P_n$;
   end;
6. Repeat step 5 for $n$ times.
7. Select TTP using $R_f$.
8. For $P_1$ to $P_n$, do
   begin
   Send $S_{nr}$ to *TTP*;
   end;
9. *TTP* decodes $S_{nr}$ using $F_n$ from $D$; and rearrange $P_nK_r$ into data blocks;
10. *TTP* computes and announces result;

## 7. ANALYSIS AND PERFORMANCE

*Extended Encrypto_Random*

**Case 1: Joint malicious conduct by certain Parties.**

The parties receive the data packets $P_nK_r$ and the function value $V_rF_n$. At a given instant, if $m$ number of parties get together, they will have the following information at their disposal: 1). $m*X_r$ number of packets, ($X_r$ = number of packets generated by a single party, as fixed by the TTP) and 2). $m$ number of encrypting functions. This information is useless for them, because they won't be having the encrypting functions of the other parties for generating the whole blocks of data. Moreover it is highly improbable that these m parties receive all the packets of a certain party, thus making the available information with them redundant.

**Case 2: The chosen TTP becomes malicious.**

The TTP receives the packets $P_nK_r$ and the function value $V_rF_n$. Also, it has the pool of random functions. At a given instant it has the set ($P_nK_r$, $V_rF_n$, pool of functions). The TTP can assemble the data packets to form whole data blocks, but in no case can relate any data block thus formed to a certain party. Further, if there are a large number of participating parties (which can be assumed to be the real world scenario, and for which our protocol is sincerely aimed at), then it becomes nearly impossible to decisively relate a particular data block to a particular party. This is also evident by the probabilistic curve, which describes an inversely parabolic path and tends to 0 as the number of parties increases.

**Case 3: Joint malicious conduct by some Party and the TTP.**

In the previous protocol there was a predetermined TTP, thus there was plausibility about some party goading the TTP to exhibit corrupt behavior, and consequently give undue advantage to such a party by revealing the regrouped data blocks. This problem is taken care of in this extended work, as the TTP is decided only at run time by the protocol. Thus it becomes virtually impossible for any party motivated by vindictive intent to collaborate with the unknown TTP for a joint malicious conduct.

## 8. PROBABILISTIC EVIDENCE

Let $P(x_n)$ be the probability that a party turns malicious, where *n* is the number of parties.

Thus, $P(x_1) = P(x_2) = P(x_3) = P(x_4) = \ldots = P(x_n) = 1/n$ (equal for all parties).

If at a given instant, a party $P_r$ (say) intends to leak the information of the other parties. The probability of the number of packets that party can decrypt is given by

= number of packets of that party / total number of packets.
$= Xr / (\sum_{r=1}^{n} Xr)$

Therefore, probability that one party exhibits malicious conduct
$= [1/n] * [Xr / (\sum_{r=1}^{n} Xr)]$

Now let's say *r* number of parties become malicious.
Thus, Probability for leak of the packets
$= [r/n] * [\sum_{r=1}^{r} Xr] / (\sum_{r=1}^{n} Xr)$

Also, let $P(t_m)$ be the probability that the chosen TTP turns malicious, where m is the number of total TTP out of which one will be chosen.

Thus, $P(t_1) = P(t_2) = P(t_3) = P(t_4) = \ldots = P(t_m) = 1/m$ (equal for all TTP).

Therefore, Total Probability for leak of the packets is given by
$= 1/m * [r/n] * [\sum_{r=1}^{r} Xr] / (\sum_{r=1}^{n} Xr)$

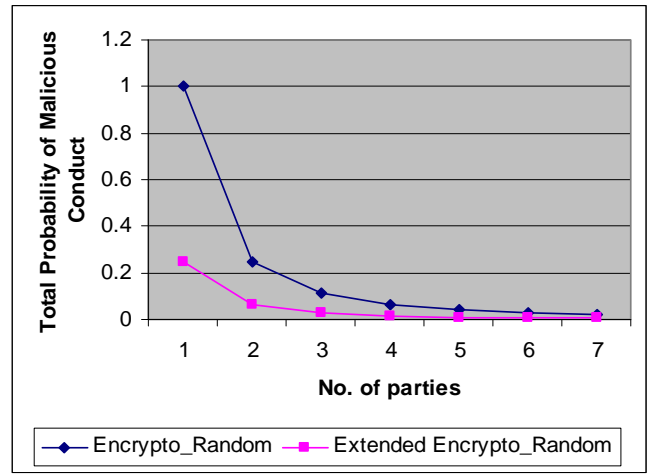

**Figure 3. Total Probability for Leak of Data**

*n* = number of parties coming together for joint computation.
*m* = number of TTP from which one is chosen.
$X_r$ = 4, fixed number of packets made by each of these *n* parties (as decided by the TTP). Here has been taken as example to construct the graph.
m = 4, fixed number of TTP taken as example to illustrate the graph.
Let's say that one party out of *n* decides to decrypt its data packets. Thus, probability that it becomes successful in exhibiting such malicious conduct is given by: $(1/m * 1/n^2)$.

This curve depicts an *inversely hyperbolic path* and substantially proves that, as the number of participating parties increases, the probability of malicious conducts and the subsequent leakage of data decreases, and ultimately tends to zero if the number of parties are numerous.

## 9. CONCLUSION

Secure Multi-Party Computation is a well researched topic. Quite a few protocols already exist, and work is going-on on another handful. Through *Extended Encrypto_Random* we have endeavored to present a concept that emphasizes the need to keep the structure of the proposed solution to the problem very forthright so as to avoid ambiguities; at the same time ensuring the security of information by taking efficient and intricate measures. The data is first distributed and then sent forward; assuring that no party becomes victim to sabotage by other parties and also that, no party gets undue privilege, as the sole responsibility of the computation process is not vested upon a single entity. The encrypted nature of data further hinders any possibility of spiteful conduct. The possibility of collaborative malefic behavior by some party and the TTP has been completely curbed by concealing the identity of the TTP until runtime. Our protocol also reduces the complexities that are encountered in three and four layer protocols.

## 10. FUTURE SCOPE

The function domain is being further developed and the transforming functions that leverage the proposed architecture in different areas are being fine-tuned. Subsequent enhancement of the protocol is expected in the sense that instead of making a single TTP responsible for calculating the result, multiple TTP are given the same set of data packets, upon which each of them will perform the same computations. Then the results of each of them are compared to decide which of them are harmonious. If more than half of the results are found synonymous, then such a result can be authenticated.

## REFERENCES

[1] W. Du, and M. J. Attalah, "Secure Multi - Party Computation Problems and Their Applications: A Review and Open Problems." *Tech Report CERIAS Tech Report 2001-51*, Centre for Education and Research in Information Assurance and Security and Department of Computer Sciences, Purdue University, West Lafayette, IN 47906, 2001.

[2] J. Vaidya, and C. Clifton, "Leveraging the Multi in Secure Multi-Party Computation." *WPES'03 October 30, 2003*, Washington DC, USA, ACM Transaction 2003, pp 120-128.

[3] M. J. Atallah and W. Du., "Secure Multi-Party Computation Geometry." *Seventh International Workshop on Algorithms and Data Structures (WADS 2001)*, Providence, Phode Island, USA, Aug 8- 10, 2001, pp 136-152

[4] W. Du, and M. J. Attalah, "Privacy-Preserving Cooperative Scientific Computations." *IN 14th IEEE Computer Security Foundation Workshop*, Nova Scotia, Canada June 11-13, 2001.

[5] U. Maurer, "The Role of Cryptography in Database Security." *SIGMOD 2004,* June, Paris, France June 13-18, 2004, pp 29-35.

[6] O. Goldreich, S. Micali, and A. Wigderson, "How to Play Any Mental Game – A Completeness Theorem for Protocols with Honesty Majority." *19th ACM Symposium on the Theory Of Computation,* 1987, pp 218-229.

[7] U. Maurer, "Secure Multi-Party Computation made Simple." *Security in Computational Network (SCN'02)*, G. Persiano (Ed.), Lecture notes in Computer Science, Springer- Verlag, Vol. 2576, 2003, pp 14-28.

[8] A. C. Yao, "Protocols for Secure Computations." *In Proc. 23rd IEEE Symposium on the Foundation of Computer Science (FOCS),* IEEE 1982, pp 160- 164.

[9] Agrawal, A. Evfimievski, and R. Srikant, "Information Sharing Across Private Databases." *SIGMOD- 2003*, SAndiego CA, June 9-11, 2003, pp 109-115.

[10] R. Canetti, "Secure and Composition of Multi-Party Cryptographic Protocols." *Journal of Cryptography,* Vol. 13, No. 1, 2000, pp143-202.

[11] B. Pfitzmann, M. Schunter, and M. Waidner, "Secure Reactive System." *IBM Research Report RZ 3206, Feb 14, 2000.

[12] J. VAidya and C. Clifton, "Privacy Preserving Association Rule Mining in Vertically Partitioned Data." *8thACM SIGKDD International Conference on Knowledge Discovery and Data Mining(2002),* pp 639- 644.

[13] Y. LIndell and B. Pinkas, "Privacy Preserving Data Mining." *Advances in Cryptology – CRYPRTO 2000,* Springer- Verlag, Aug 20-24, 2000, pp 36-54.

[14] H. Kargupta, B. Park, D. Herchberher, and E. Johnson, "Collective Data Mining: A New Perspective Toward Distributed Data MInig Book, Philip Chan, AAA Press, 1999.

[15] Ivan Demgard and Yuvai Ishai, "Constant-Round Multi-Party Computation using a Black-Box Pseudo Random Generator." Aug 10, 2005.

[16] A C Yao, "How to generate and Exchange Secrets." April 16, 2004.

[17] Yehuda Lindell, "Parallel Coin-Tossing and ConstantRound Secure Two-Party Computation."

[18] D. K. Mishra, M. Chandwani, "Secure Multi-Party Computation Protocol using Ambiguous Identity," accepted for the publication in Journal.

[19] D. K. Mishra, M. Chandwani, "Extended Protocol for Secure Multi-party Computation using Ambiguous Identity". *WSEAS Transaction on Computer Research, Vol 2, issue 2, Feb 2007.*

[20] Neha Koria, Nikhil Kapoor, Ravish Bahety, D K Mishra, "Complexity Minimization in Secure Multi-Party Computation for Preserving Privacy during Data Mining." *Computing for National Development, INDIACOM'09, 3rd National Conference, 26-27 Feb 2009.*

## AUTHORS PROFILE

*Dr. Durgesh Kumar Mishra*

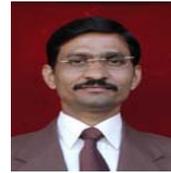

**Biography:** Dr. Durgesh Kumar Mishra has received M.Tech. degree in Computer Science from DAVV, Indore in 1994 and PhD degree in Computer Engineering in 2008. Presently he is working as Professor (CSE) and Dean (R&D) in Acropolis Institute of Technology and Research, Indore, MP, India. He is having around 20 Yrs of teaching experience and more than 5 Yrs of research experience. He has completed his research work with Dr. M. Chandwani, Director, IET-DAVV Indore, MP, India in Secure Multi- Party Computation. He has published more than 60 papers in refereed International/National Journal and Conference including IEEE, ACM etc. He is a senior member of IEEE and Secretary of IEEE MP-Subsection under the Bombay Section, India. Dr. Mishra has delivered his tutorials in IEEE International conferences in India as well as other countries also. He is also the programme committee member of several International conferences. He visited and delivered his invited talk in Taiwan, Bangladesh, USA, UK etc in Secure Multi-Party Computation of Information Security. He is an author of one book also. He is also the reviewer of tree International Journal of Information Security. He is a Chief Editor of Journal of Technology and Engineering Sciences. He has been a consultant to industries and Government organization like Sale tax and Labor Department of Government of Madhya Pradesh, India.

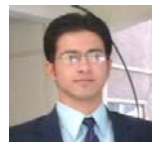

Nikhil Kapoor (India), He has done BE in Computer Engineering and Science in 2009. He has published paper in referred International/National Journal and Conferences.

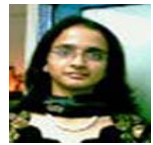

Neha Koria (India), She has done BE in Computer Engineering and Science in 2009. He has published paper in referred International/National Journal and Conferences.

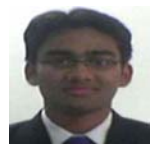

Ravish Bahety (India), He has done BE in Computer Engineering and Science in 2009. He has published paper in referred International/National Journal and Conferences.